\documentclass[floats,floatfix,showpacs,preprintnumbers,amssymb,prd,twocolumn,superscriptaddress,nofootinbib,nolongbibliography,reprint]{revtex4-1}

\usepackage{amssymb,amsmath,verbatim,mathtools,needspace,enumitem,etoolbox,graphicx,physics,microtype,afterpage,xspace,tabularx,lmodern,multirow,boldline, makecell, booktabs,mathrsfs}
\usepackage{gensymb}
\usepackage{appendix}
\usepackage{tensor}
\usepackage{array}
\usepackage[normalem]{ulem}
\usepackage[dvipsnames, usenames]{xcolor}
\usepackage{xr-hyper}
\definecolor{linkcolor}{rgb}{0.0,0.3,0.5}
\usepackage[unicode, colorlinks=true, linkcolor=linkcolor, citecolor=linkcolor, filecolor=linkcolor, urlcolor=linkcolor, linktocpage, breaklinks]{hyperref}
\usepackage[all]{hypcap}
\usepackage[T1]{fontenc}
\usepackage[utf8]{inputenc}
\hypersetup{colorlinks=true,citecolor=magenta,linkcolor=violet,urlcolor=magenta}
\usepackage{multirow}

\usepackage{xargs}
\usepackage{multirow}

\newcommand{\sch}{Schwarzschild }
\newcommand{\jj}[6]{\begin{pmatrix}#1 & #2 & #3 \\ #4 & #5 & #6 \end{pmatrix}}

\makeatletter
\newcommand*{\addFileDependency}[1]{
  \typeout{(#1)}
  \@addtofilelist{#1}
  \IfFileExists{#1}{}{\typeout{No file #1.}}
}
\makeatother

\begin{document}
\title{Ringdown nonlinearities in the eikonal regime}

\author{Bruno Bucciotti}
\affiliation{Scuola Normale Superiore, Piazza dei Cavalieri 7, 56126 Pisa, Italy}
\affiliation{INFN, Sezione di Pisa, Largo B. Pontecorvo, 3, 56127 Pisa, Italy}
\author{Vitor Cardoso} 
\affiliation{Center of Gravity, Niels Bohr Institute, Blegdamsvej 17, 2100 Copenhagen, Denmark}
\affiliation{CENTRA, Departamento de F\'{\i}sica, Instituto Superior T\'ecnico--IST, Universidade de Lisboa--UL, Avenida Rovisco Pais 1, 1049-001 Lisboa, Portugal}
\author{Adrien Kuntz}
\affiliation{CENTRA, Departamento de F\'{\i}sica, Instituto Superior T\'ecnico--IST, Universidade de Lisboa--UL, Avenida Rovisco Pais 1, 1049-001 Lisboa, Portugal}
\author{David Pereñiguez}
\affiliation{Center of Gravity, Niels Bohr Institute, Blegdamsvej 17, 2100 Copenhagen, Denmark}
\author{Jaime Redondo-Yuste}
\affiliation{Center of Gravity, Niels Bohr Institute, Blegdamsvej 17, 2100 Copenhagen, Denmark}

\date{\today}

\begin{abstract}
The eikonal limit of black hole quasinormal modes (the large multipole limit $\ell\gg1$) can be realized geometrically as a next-to-leading order solution to the geometric optics approximation, and also as linear fluctuations about the Penrose limit plane wave adapted to the light ring. Extending this interpretation beyond the linear order in perturbation theory requires a robust understanding of quadratic quasinormal modes for large values of $\ell$. We analyze numerically the relative excitation of quadratic to linear quasinormal modes of Schwarzschild black holes, with two independent methods. Our results suggest that the ratio of quadratic to linear amplitudes for the $\ell \times \ell \rightarrow 2\ell$ channel converges towards a finite value for large $\ell$, in sharp contrast with a recent proposal inspired by the Penrose limit perspective. 
On the other hand, the $2 \times \ell \rightarrow \ell+2$ channel seems to have a linearly growing ratio. Nevertheless, we show that there is no breakdown of black hole perturbation theory for physically realistic initial data.
\end{abstract}

\maketitle

\noindent {\bf \em Introduction.}
Black hole (BH) spectroscopy aims to understand the characteristic modes, or quasinormal modes (QNMs), of BHs. Most processes involving BHs excite their QNMs, and the consequent gravitational-wave signal carries precious information about the BH mass and spin, its environment and the underlying theory of gravity~\cite{Kokkotas:1999bd,Berti:2009kk,Cardoso:2019rvt,Baibhav:2023clw}. In the high-frequency regime or eikonal limit, QNMs are described by a large angular index $\ell \gg 1$ in the spherical harmonics expansion. This limit can be associated to a null geodesic congruence centered at the light ring (LR) with an expansion determined by Lyapunov's exponent \cite{Mashhoon:1985cya, Schutz:1985km, Cardoso:2008bp}. It can also be understood as fluctuations about the Penrose limit adapted to the LR.\footnote{The Penrose limit adapted to a null geodesic $\gamma$ is a Ricci-flat $pp$-wave that can be seen as a zoom into the spacetime near $\gamma$ \cite{Penrose1976AnySH}.} The latter analogy was made precise in Ref.~\cite{Fransen:2023eqj} where, in particular, the boundary conditions for the Penrose plane wave fluctuation corresponding to large-$\ell$ QNMs were identified. This correspondence was established for linear scalar field fluctuations and was shown to be equivalent to the next-to-leading geometric optics approximation. Recent work in this direction has been carried out in Refs.~\cite{Giataganas:2024hil,Kapec:2024lnr}.

On the other hand, fully nonlinear numerical simulations have shown clear signs of so-called quadratic QNMs (QQNMs)~\cite{Ma:2022wpv,London:2014cma,Cheung:2022rbm,Mitman:2022qdl,Cheung:2023vki,Khera:2023oyf,Zhu:2024rej,Redondo-Yuste:2023seq}, a consequence of the nonlinear nature of general relativity. Indeed, at second order in perturbation theory, two linear QNMs can combine due to mode coupling. These extra modes are exciting because they probe general relativity deeper in its nonlinear regime, since not only their frequencies, but also their amplitudes are determined only by the properties of the remnant BH, and the amplitudes of the linear modes themselves~\cite{Nakano:2007cj,Bucciotti:2024zyp,Bucciotti:2024jrv,Ma:2024qcv,Khera:2024yrk,Bourg:2024jme}.
Is there a similar correspondence between QQNMs and properties of the LR?

Answering this question requires, as a first step, a robust evaluation of the relative amplitudes (or \textit{amplitude ratio} $\mathcal{R}$) between quadratic to linear fluctuations in the eikonal regime. Define 
\begin{equation}
\mathcal{R}^{\ell}_{\ell_1 m_1\times \ell_2 m_2}\equiv \frac{A^{(2)}_{\ell m}}{A^{(1)}_{\ell_1 m_1}A^{(1)}_{\ell_2 m_2}}\,,\;m=m_1+m_2\,,\label{ratio_def}
\end{equation}
where $A^{(1)},A^{(2)}$ are projected components of the linear and quadratic amplitudes of the gravitational-wave strain measured at large distances, or directly at future null infinity~\cite{Cheung:2022rbm}. These are projected onto spherical harmonics and each carry different angular index $\ell' m'$. We will always be dealing with the fundamental mode in this work, and therefore do not use any overtone label, unless otherwise stated.

A given QQNM depends on the properties of its parent linear modes and of the angular indices $\ell, m$ of the quadratic mode itself~\cite{Bucciotti:2024jrv}. This is why (following standard notation) we will denote a given QQNM by $(\ell_1, m_1) \times (\ell_2, m_2) \rightarrow (\ell, m)$.~\footnote{We focus on modes with a positive real part of the frequency.} 
Here, we obtain their amplitude for \sch BHs following two independent approaches, presented in Refs.~\cite{Redondo-Yuste:2023seq,Bucciotti:2024zyp,Bucciotti:2024jrv}.

\noindent {\bf \em Numerical procedure.}
%
We focus on initial conditions that respect equatorial symmetry, such as those excited in nonprecessing BH mergers. In that case, the amplitude of ``mirror'' modes is fixed relative to the amplitude of positive frequency QNMs, and the ratio of quadratic to linear amplitudes is characterized by a single number~\cite{Bourg:2024jme,Bucciotti:2024jrv,Khera:2024yrk}.
We obtain this single number by two different methods that we now describe. We also compare our results to a recent proposal~\cite{Kehagias:2024sgh} where the amplitude ratio is computed using arguments based on a generalization of the Penrose limit. More precisely, a generalization of previous results, which we discuss below, would predict the following nonlinear ratio for the $(\ell_1, \ell_1) \times (\ell_2, \ell_2) \rightarrow  (\ell_1+\ell_2, \ell_1+\ell_2)$ QQNM, in units where the BH mass is $M=1$:
\begin{equation}\label{eq:R_l1_l2}
\mathcal{R}_{\ell_1 \ell_1 \times\ell_2 \ell_2}^{\ell_1+\ell_2} =  \frac{\ell_1\ell_2}{27} \, .
\end{equation}
Our first method relies on the numerical evolution of the second order Teukolsky equation~\cite{Ripley:2020xby}, followed by the extraction of the linear fundamental mode and the quadratic mode from the waveform, by minimizing the mismatch between the numerical waveform and a model based on a superposition of damped sinusoids, as implemented in \texttt{Jaxqualin}~\cite{Cheung:2023vki}. Both steps introduce different sources of error. We estimate the uncertainty by computing how the recovered amplitudes vary when the starting time of the fit changes. We highlight that this ``scattering approach'' seem to always overestimate the amplitude ratio, perhaps due to an assimilation of part of the overtone content of the signal in the amplitude of the QQNMs. Since this method is based on a time domain scattering experiment, it has been observed that the amplitudes can vary mildly depending on the initial data, in particular, depending on the wavelength of the initial fluctuation~\cite{Redondo-Yuste:2023seq, Zhu:2024rej}. Whether this initial data dependence is physical (due to, e.g., coupling between QNMs and tail or burst contributions), or simply due to systematics of the fitting procedure, is not yet clear, but in any case, it provides an additional source of unmodeled uncertainty.
This method requires increasing resolution for higher values of $\ell$, therefore, in order to ensure that the numerical evolution is converging we restrict ourselves to $\max(\ell_1,\ell_2)\leq 5$.

The second method (labeled ``Leaver'' onward) is based on a solution of the Regge-Wheeler-Zerilli equations at second order with QNM boundary conditions using the Leaver algorithm~\cite{Bucciotti:2024zyp,Bucciotti:2024jrv}. The solutions are obtained as an infinite series whose coefficients can be computed with arbitrary precision in \texttt{Mathematica}. This means that the accuracy of the method is very high, so that the uncertainty on the quadratic ratio is well below the first method.

\noindent {\bf \em Results.}
%
\begin{table}[h!]
\setlength{\tabcolsep}{6pt} 
\renewcommand{\arraystretch}{0.7} 
\begin{tabular}{lllll}
\hline\hline
 &  $\ell=2$ & $\ell=3$ &  $\ell=4$ & $\ell=5$ \\ \hline
Ref.~\cite{Kehagias:2024sgh} & $0.148$ & $0.333$ & $0.592$ & $0.926$ \\
Scattering~\cite{Redondo-Yuste:2023seq} & $0.17(1)$ & $0.239(5)$ & $0.28(2)$ & $0.25(2)$ \\
Leaver~\cite{Bucciotti:2024jrv} & $0.154$ & $0.231$ & $0.238$ & $0.230$ \\ \hline\hline
\end{tabular}
\caption{
Summary of the results for the nonlinear ratio $\mathcal{R}_{\ell \ell \times\ell \ell}^{2 \ell}$ of the $(\ell, \ell) \times (\ell, \ell) \rightarrow  (2\ell, 2 \ell)$ QQNM for $\ell=2,3,4,5$ according to Ref.~\cite{Kehagias:2024sgh} (first row), based on our scattering experiments, and using the Leaver algorithm. In the second row, the digit in parentheses indicates the statistical uncertainty in the last digit.\label{tab:theTable}}
\end{table}
\begin{figure}
    \centering
    \includegraphics[width=\columnwidth]{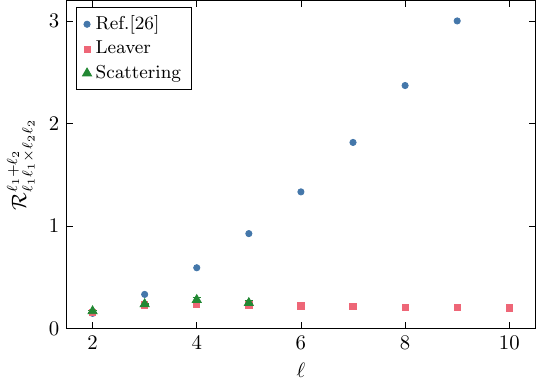}
    \caption{Ratio $\mathcal{R}_{\ell \ell \times\ell \ell}^{2 \ell}$ obtained using the formula based on the Penrose limit in Eq.~\eqref{eq:R_l1_l2} (blue circles), based on a calculation using Leaver's method (red squares), and extracted from numerical scattering experiments (green triangles), for different values of $\ell$. While both scattering and Leaver methods agree up to their statistical uncertainty, and seem to indicate a value that asymptotes to a constant, the result of Ref.~\cite{Kehagias:2024sgh} behaves completely differently as $\ell\to\infty$. The agreement for $\ell=2$ seems, then, purely coincidental. }
    \label{fig:Ratio}
\end{figure}
There are different possibilities when taking the eikonal limit on a QQNM, due to the relative scaling of $\ell_1$, $\ell_2$, $m_1$, $m_2$ and $\ell$. We first examine the nonlinear ratio $\mathcal{R}_{\ell \ell \times\ell \ell}^{2 \ell}$ of the $(\ell, \ell) \times (\ell, \ell) \rightarrow  (2\ell, 2 \ell)$ QQNM. 
Our results are summarized in Table~\ref{tab:theTable} and Fig.~\ref{fig:Ratio}. The numerical agreement between our methods and the prediction in Ref.~\cite{Kehagias:2024sgh} for $\ell=2$ is coincidental. Our results and the predictions of Ref.~\cite{Kehagias:2024sgh} diverge for $\ell>2$. In particular, the scattering and Leaver methods indicate that the nonlinear ratio $\mathcal{R}_{\ell \ell \times\ell \ell}^{2 \ell}$ either saturates to approximately $0.2$ or has only a mild dependence on $\ell$ in the eikonal regime, whereas Ref.~\cite{Kehagias:2024sgh} would predict a rapid growth. This disagreement indicates that the some of the arguments in the calculation of Ref.~\cite{Kehagias:2024sgh} are not valid (we expand on this below). 

\begin{figure}
    \centering
    \includegraphics[width=\columnwidth]{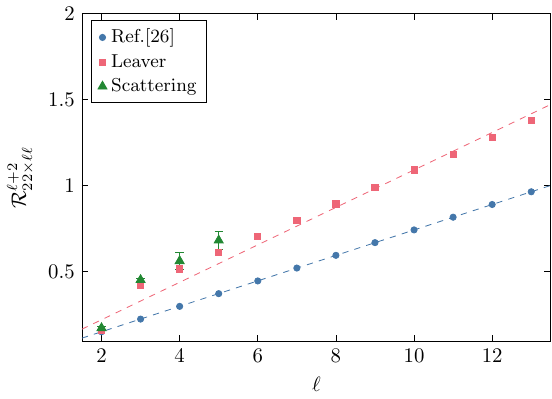}
    \caption{Ratio $\mathcal{R}_{2 2 \times\ell \ell}^{\ell+2}$ obtained using Eq.~\eqref{eq:R_l1_l2} (blue circles), using the Leaver method (red squares), and extracted from numerical scattering experiments (green triangles), for different values of $\ell$. All methods agree at predicting a linear growth $\mathcal{R}_{2 2 \times\ell \ell}^{\ell+2}\sim \ell$, but Ref.~\cite{Kehagias:2024sgh} fails to accurately predict the slope of the line. The dashed lines show the best fit to a linear polynomial in $\ell$ removing the first few points, which lead to $\mathcal{R}_{2 2 \times\ell \ell}^{\ell+2} \sim 0.11 \ell$ based on Leaver's method, as opposed to $\mathcal{R}_{2 2 \times\ell \ell}^{\ell+2} = 0.074 \ell$ based on Eq.~\eqref{eq:R_l1_l2}. }
    \label{fig:ratio_l_2}
\end{figure}

Additionally, we have studied different channels, beyond the $(\ell,\ell)\times(\ell,\ell) \to (2\ell,2\ell)$ discussed above (while always restricting to couplings between fundamental modes). In particular, we turn now our attention to modes that are excited by the $(2,2)$ mode in combination with another mode $(\ell,\ell)$, as extracted in the $(\ell+2,\ell+2)$ angular sector. The values that we obtain for $\mathcal{R}_{2 2 \times\ell \ell}^{\ell+2}$ are 
summarized in Table~\ref{tab:ell_vs_2} and in Fig.~\ref{fig:ratio_l_2}. A very interesting feature emerges: the ratio seems to grow linearly with $\ell$. We obtain $\mathcal{R}_{2 2 \times\ell \ell}^{\ell+2} \approx 0.11\ell \neq 2\ell/27$, therefore also in disagreement with Eq.~\eqref{eq:R_l1_l2}, as apparent in Fig.~\ref{fig:ratio_l_2}. We comment on this property below.

\begin{table}[h!]
\setlength{\tabcolsep}{6pt} 
\renewcommand{\arraystretch}{0.7} 
\begin{tabular}{llll}
\hline\hline
 &  $\ell=3$ &  $\ell=4$ & $\ell=5$ \\ \hline
Ref.~\cite{Kehagias:2024sgh}& 0$.222$  & $0.296$ & $0.370$ \\
Scattering & $0.45(1)$ & $0.56(5)$ & $0.68(5)$ \\
Leaver & $0.417$ & $0.514$ & $0.608$ \\ \hline\hline
\end{tabular}
\caption{Summary of the results for the nonlinear ratio $\mathcal{R}_{2 2 \times\ell \ell}^{\ell+2}$ for $\ell=3,4,5$ according to Ref.~\cite{Kehagias:2024sgh}, based on scattering experiments, and Leaver's method. In the second row, the digit in parentheses indicates the statistical uncertainty in the last digit. \label{tab:ell_vs_2}}
\end{table}
Finally, we comment on the $m$ dependence of our results. All quadratic amplitudes we reported were always computed assuming $m_1=\ell_1,\,m_2=\ell_2$, but we can easily extend these results due to the rotational symmetry enjoyed by Schwarzschild BHs, which implies that the $m$ dependence of QQNMs is fully captured by a $3j$ symbol,
\begin{equation}
\label{eq:3j_general}
    \mathcal{R}_{\ell_1 m_1\times \ell_2 m_2}^{\ell} \propto \jj{\ell_1}{\ell_2}{\ell}{m_1}{m_2}{-m_1-m_2} \, .
\end{equation}
This property, along with an estimate of the $\ell\rightarrow\infty$ limit of $3j$ symbols discussed in Appendix~\ref{app:3j}, allows us to rescale our previous results giving
\begin{equation}
\label{eq:3j_2l}
    \mathcal{R}_{\ell m_1\times \ell m_2}^{2\ell} \simeq 
    (-1)^{m_1+m_2}\left(\frac{2}{\ell \pi }\right)^{1/4} e^{-\frac{(m_1-m_2)^2}{4\ell}}
    \mathcal{R}_{\ell \ell \times \ell \ell}^{2\ell}\,,
\end{equation}
valid when $m_1,m_2\lesssim \sqrt{\ell}$. This limitation explains the apparent discrepancy if we tried to set $m_1=m_2=\ell$. 

For $\ell\times 2\rightarrow\ell+2$, we have instead
\begin{equation}
\label{eq:3j_l2}
    \mathcal{R}_{\ell m_1\times 2 2}^{\ell+2} \simeq 
    \frac{(-1)^{\ell +m_1}}{4} \left(1+\frac{m_1}{\ell }\right)^2
    \mathcal{R}_{\ell \ell \times 2 2}^{\ell+2}\,,
\end{equation}
uniformly in $|m_1|\le\ell$.

\noindent {\bf \em Eikonal QNMs from the Penrose limit.}
Our results are in clear tension with the claims in Ref.~\cite{Kehagias:2024sgh}. We now briefly revisit their construction to find the nonlinear ratio based on the Penrose limit along the LR. Following the spirit of their argument, we also extend the analysis to account for modes with $\ell_1 \neq \ell_2$, as well as overtones. 
As we highlight, several assumptions require further scrutiny and ultimately fail (as our results demonstrate).
For simplicity we restrict our discussion to fluctuations about a Schwarzschild BH, but the arguments readily extend to Kerr. The Penrose limit of a Schwarzschild BH along the LR, at $r_{0}=3M$ and $\theta_{0}=\pi/2$, is a Ricci-flat plane wave~\cite{Blau:2003dz}
\begin{equation}\label{SchPP}
    ds^{2}=2 du dv +\left(\frac{z^{2}+\bar{z}^{2}}{3 M^{2}}\right)du^{2}-2d \bar{z}dz\, . 
\end{equation}
Here, the LR lies at $v=z=0$, while $\delta r \equiv (z+\bar{z})/\sqrt{6}$ and $\delta \theta\equiv i (\bar{z}-z)/r_{0}\sqrt{2}$ measure the deviation away from it along the radial and axial directions. This is a Petrov-type $N$ space, $\boldsymbol{l}=\partial_{v}$ being the principal null direction. Consequently, gravitational fluctuations can be generated from a Hertz potential $\Phi$~\cite{Wald:1978vm}. By definition it satisfies $\square \Phi =0$ on the background in Eq.~\eqref{SchPP}. Solutions that are also eigenfunctions of $\partial_{v}$ and $\partial_{u}$ read 
\begin{equation}\label{Hertz}
\Phi = A \, e^{i\left( P_{u} u +  P_{v} v\right)}\chi(z,\bar{z})\, ,
\end{equation}
for some constants $P_{u}$ and $P_{v}$ and some function $\chi(z,\bar{z})$ of $z,\bar{z}$ only. These solutions, subject to the boundary conditions derived in Ref.~\cite{Fransen:2023eqj}, yield quantized values for $P_{u,v}$,
\begin{equation}
\begin{aligned}\label{Ps}
P_{v}&=\frac{m}{3 \sqrt{3}M}\, , \\
P_{u}&=\frac{1}{\sqrt{3} M }\left[m-\ell-1/2+i\left(1/2+n\right)\right] \, ,
\end{aligned}
\end{equation}
where $n,\ell,m$ are integers. From these values, and relating the plane wave coordinates of Eq.~\eqref{SchPP} to the original Schwarzschild coordinates, Ref.~\cite{Fransen:2023eqj} obtained the well-known large-$\ell$ or eikonal limit of QNM frequencies, where $n,\ell,m$ in Eq.~\eqref{Ps} correspond to the overtone and harmonic quantum numbers, respectively. We remark here that the identification between perturbations on the Penrose LR wave and eikonal QNMs holds, in principle, only at the level of the quasinormal frequencies, and does not necessarily capture the behaviour of amplitudes.

Reference~\cite{Kehagias:2024sgh} proposes to extend this picture to QQNMs as follows. Consider a first-order metric fluctuation on a Penrose LR wave, which upon acting with reconstruction operators on Eq.~\eqref{Hertz} is given by $h_{ab}\equiv \gamma_{ab}+c.c.$, with 
\begin{equation}\label{pert}
    \gamma=\Phi_{\bar{z}\bar{z}}du^{2}+2\Phi_{v\bar{z}}du dz+\Phi_{vv}dz^{2}\, .
\end{equation}
Then, in a certain limit that focuses further on the LR at $v=z=0$, the perturbation in Eq.~\eqref{pert} becomes a function of $u$ only, as dictated by the formal substitution given in Eq.~(8) of Ref.~\cite{Kehagias:2024sgh}, yielding
\begin{equation}\label{pert2}
    \gamma=A e^{i P_{u} u}\left(2\alpha_{\bar{z}\bar{z}}du^{2}-P_{v}^{2}dz^{2}\right)\, ,
\end{equation}
for some constant $\alpha_{\bar{z}\bar{z}}$. We note that this limit cannot correspond to taking a solution $\Phi(u,\lambda^{2}v,\lambda z,\lambda \bar{z})$, constructed out of Eq.~\eqref{Hertz}, and sending $\lambda\to0$ as the authors seem to suggest (that would result in a vanishing metric fluctuation, instead of Eq.~\eqref{pert2}). In fact, Eq.~\eqref{pert2} does not even satisfy the linearized Einstein equations, which leads us to conclude it only corresponds to the metric in Eq.~\eqref{pert} evaluated exactly at the LR. Next, Ref.~\cite{Kehagias:2024sgh} argues that the limit fluctuation in Eq.~\eqref{pert2} is arbitrarily close to the action of a large diffeomorphism which, close to the LR, can be absorbed in a background redefinition by introducing a new coordinate $v'=v+g(u)$, with
\begin{equation}\label{g}
g(u)=-i\frac{\alpha_{\bar{z}\bar{z}}}{P_{u}}A e^{i P_{u}u}+\mathrm{c.c.}\, .
\end{equation}
Crucially, the function $g(u)$ is proportional to the amplitude $A$ of Eq.~\eqref{pert2}. 

Finally, one takes a linear fluctuation of the redefined background and, from that, reads off the quadratic to linear amplitude ratios. Here, we follow the logic of Ref.~\cite{Kehagias:2024sgh} and generalize the argument to account for arbitrary pairs of linear fundamental modes, with $\ell=\{\ell_1,\ell_2\}$ and $m_1=\ell_1$ and $m_2=\ell_2$, with amplitudes $A_1$ and $A_2$ (our results reduce to those in Ref.~\cite{Kehagias:2024sgh} for $\ell_{1}=\ell_{2}$). According to Ref.~\cite{Kehagias:2024sgh}, a superposition of two linear fundamental modes should correspond to a superposition of two large diffeomorphisms, generated by a coordinate transformation $v^\prime = v+g_1(u)+g_2(u)$, where the $g_i$ are given by Eq.~\eqref{g} (we recall that the argument until this point is fully linear). Hence, the ansatz for the fluctuation is
\begin{equation}
        \Phi = \frac{1}{2}\Bigl(A_1 e^{iP_u^{(1)}u+iP_v^{(1)}v^\prime}+A_2 e^{iP_u^{(2)}u+iP_v^{(2)}v^\prime}\Bigr) \, , 
\end{equation}
which contains both the first and second order contributions. Expanding $v'$ we find 
\begin{equation}
        \Phi \supseteq \frac{A_1A_2}{2} e^{i(P_u^{(1)}+P_u^{(2)})u}\Bigl(\frac{P^{(1)}_v \alpha^{(2)}_{\bar{z}\bar{z}}}{iP^{(2)}_u}+\frac{P^{(2)}_v \alpha^{(1)}_{\bar{z}\bar{z}}}{iP^{(1)}_u}\Bigr) \, , 
\end{equation}
where we only wrote the crossed term proportional to $A_1A_2$. Therefore, the nonlinear ratio $\mathcal{R}^{\ell_1+\ell_2}_{\ell_1 \ell_1\times \ell_2 \ell_2}$ is given by
\begin{equation}
   \mathcal{R}^{\ell_1+\ell_2}_{\ell_1 \ell_1\times \ell_2 \ell_2} = \left\lvert \frac{P^{(1)}_uP^{(1)}_v\alpha^{(2)}_{\bar{z}\bar{z}}+P^{(2)}_uP^{(2)}_v\alpha^{(1)}_{\bar{z}\bar{z}}}{P^{(1)}_uP^{(2)}_u}\right\rvert = \frac{\ell_1\ell_2}{27} \,,
\end{equation}
where in the last equality we have focused on fundamental modes. If $\ell_1=\ell_2\equiv \ell$, one recovers the result reported in Ref.~\cite{Kehagias:2024sgh}, showing a quadratic scaling in the angular number. It is not obvious that the result obtained this way should reproduce the eikonal limit of QQNM amplitudes (it does not). A particularly clear reason, though not the unique one, is that nonlinear ratios depend on the spatial coordinate where they are evaluated. For us, $\mathcal{R}^{\ell}_{\ell_1 m_1\times \ell_2 m_2}$ are defined at infinity, and this has not been accounted for in the argument above based on Ref.~\cite{Kehagias:2024sgh}. 

\noindent {\bf \em High frequency limit.}
Reference~\cite{Kehagias:2024sgh} suggested a divergence in the nonlinear ratio in the high-frequency (or eikonal) regime. Although we have now shown these results to disagree with our numerical calculations, and pointed out plausible flaws in their derivation, we still observe a divergent ratio $\mathcal{R}^{\ell+2}_{\ell m_1 \times 22} \sim \ell$. Naturally one could worry that such a divergence would imply that perturbation theory is not valid at high enough frequencies, since the second order modes could in principle have a much larger amplitude than the leading order modes. Here we argue that this can not be the case, at least for physically reasonable initial conditions. 

The perturbative expansion on which linear and quadratic modes is based is not necessarily guaranteed to converge. It can break down if certain resonances between modes are present~\cite{Bizon:2011gg, Balasubramanian:2014cja, Bizon:2015pfa,Yang:2014tla, Iuliano:2024ogr}.
Upon direct examination, one can conclude that such resonances do not occur, at least for the lower $\ell$ modes, with low overtone number, for Schwarzschild. 

An exact resonance can be understood as a divergent ratio for certain quadratic or cubic nonlinearities. However, a breakdown of perturbation theory can also occur if the nonlinear ratio describing the coupling between several incoming modes with wave number $k_i$, and resulting in a mode with wave number $k \gtrsim k_i$, diverges as $k\to\infty$. Let us be more precise. Suppose that the nonlinearity is quadratic, hence leading to a three wave interaction~\cite{nazarenko2011wave}, or equivalently, a vertex in a Feynman diagram with valence $3$. Assume that the coupling coefficient in that vertex, i.e., the nonlinear ratio between the amplitude of the outgoing (quadratic) mode, and the incoming (linear) modes, scales as $\mathcal{R}(k) \sim |k|^\alpha$ for some positive power $\alpha$. If the initial conditions are such that $A_k \sim k^{-(\alpha-\delta)}$ for any positive value of $\delta$, perturbation theory will break down for sufficiently high wave number, regardless of how small the initial perturbations are. 

How does this translate to our original problem? For a Schwarzschild BH, the angular number $\ell$ plays the role of the wave number $k$, and the leading order nonlinearity is quadratic. The amplitudes that are important in order to describe a breakdown of perturbation theory are the amplitudes of the metric perturbation, $A_\ell$. Let us examine the $\ell\times 2$ coupling discussed above, where we observe precisely a linear growth with wave number $\mathcal{R}(\ell) \propto \ell$. Thus, if we could excite the BH with initial data such that $\ell A_\ell \to \infty$ as $\ell \to \infty$, perturbation theory would break down. We argue that such an initial spectrum is not physical. Indeed, consider a point particle being shot radially towards a Schwarzschild BH. If it is shot exactly towards the center, in the ultrarelativistic regime, this leads to a spectrum $A_\ell \sim \ell^{-2}$~\cite{Sperhake:2008ga}. Thus, this is clearly not enough to trigger an instability, since the high-frequency modes are not sufficiently excited. We can imagine an enhancement of this high-frequency content by increasing the impact parameter with which the particle is thrown into the BH, until reaching the critical impact parameter. Close to that value, we would expect the spectrum to be bounded by that of a massless particle orbiting the LR. In that case, the total energy diverges~\cite{Barausse:2021xes}, but from the energy flux computed in a single orbit, we can estimate~$A_\ell \sim \ell^{-3/2}$. Once again, this does not meet the condition we established to trigger a breakdown of perturbation theory. 

A possible interpretation of this impossibility is that the nonlinear ratio in this case scales as $\mathcal{R}(\ell) \sim \omega(\ell)$, since, in the eikonal regime, $\omega_\ell = \Omega \ell +\mathscr{O}(\ell^0)$, where $\Omega$ is the orbital frequency of the LR. Thus, a breakdown of perturbation theory only occurs if the initial spectrum satisfies $A_\ell \omega_\ell \to \infty$. The Rayleigh-Jeans spectrum, typically associated with thermal states in hydrodynamics, is given by $A_\ell \omega_\ell = \mathrm{const}.$~\cite{nazarenko2011wave}. Therefore, not even a thermal state would be enough to trigger an instability. 

\noindent {\bf \em Final words.}
In this note we have studied the behavior of the ratio of quadratic QNMs in the limit of large angular number $\ell_i,\,\ell$. At large $\ell$, we find that the relative amplitudes of the $(\ell, \ell, 0) \times (\ell, \ell, 0) \rightarrow  (2\ell, 2 \ell)$ channel converge to a finite value. This result is in tension with a recent proposal predicting a $\sim \ell^{2}$ scaling for the relative amplitudes~\cite{Kehagias:2024sgh}, in particular the expression of Eq.~\eqref{eq:R_l1_l2}. 

In addition, we also evaluate the amplitude ratio $\mathcal{R}_{22 \times\ell \ell}^{\ell+2}$ of the $(2, 2, 0) \times (\ell, \ell, 0) \rightarrow  (\ell+2, \ell+2)$ QQNM. We find the surprising result that $\mathcal{R}_{22 \times\ell \ell}^{\ell+2}$ is linearly growing with $\ell$, although with a different slope from the one in Eq.~\eqref{eq:R_l1_l2}. This might seem to imply a breakdown of BH perturbation theory for some specific initial conditions. However, as argued above, these initial conditions need to be fine tuned and do not seem to be excited by simple physical processes, such as high-energy sources plunging into a BH. This opens a window towards exploring analytical methods to understand the high-frequency limit of the nonlinear ratio that excites quadratic modes in more generic configurations, as well as for Kerr BHs. 
In particular, we conjecture that initial data which can be excited through physical processes can never trigger a breakdown of perturbation theory for subextremal Kerr BHs, up to second order effects. Based on the discussion above, this would be the case if the high-frequency behavior of the quadratic to linear ratio satisfies $\mathcal{R}(\omega)/ \omega^{1+\delta} \to 0$, for any positive $\delta$. The nonlinear stability of Schwarzschild~\cite{Dafermos:2021cbw} seems to impose on us this condition.

\noindent {\bf \em Acknowledgments.} 
We thank Kwinten Fransen for valuable discussions.
We acknowledge support by VILLUM Foundation (grant no. VIL37766) and the DNRF Chair program (grant no. DNRF162) by the Danish National Research Foundation.
V.C.\ is a Villum Investigator and a DNRF Chair.  
V.C. acknowledges financial support provided under the European Union’s H2020 ERC Advanced Grant “Black holes: gravitational engines of discovery” grant agreement no. Gravitas–101052587. 
Views and opinions expressed are however those of the author only and do not necessarily reflect those of the European Union or the European Research Council. Neither the European Union nor the granting authority can be held responsible for them.
This project has received funding from the European Union's Horizon 2020 research and innovation programme under the Marie Sklodowska-Curie grant agreement No 101007855 and No 101131233.
A.K. acknowledges funding from the FCT project ``Gravitational waves as a new probe of fundamental physics and astrophysics'' grant agreement 2023.07357.CEECIND/CP2830/CT0003.
The Tycho supercomputer hosted at the SCIENCE HPC center at the University of Copenhagen was used for supporting this work.  

\bibliography{ref}
\appendix 
\onecolumngrid

\section{Eikonal limit of $3j$ symbols}
\label{app:3j}
 
In this appendix we discuss in detail the eikonal limit of $3j$ symbols, which characterize the $m$ dependence of the nonlinear ratio, in spherical symmetry. The main result is due to Ponzano and Regge \cite{ponzanoregge}, but we prefer to derive simpler (albeit not as general) expressions from first principles.

In the following, we take $\ell\rightarrow\infty$ and distinguish between small and large $m$'s, while all other parameters remain finite. First, we have
\begin{align}
    \jj{\ell}{\ell-\xi}{2\ell-\xi-\lambda}{\ell}{\ell-\xi-\lambda}{-(2\ell-\xi-\lambda)}\simeq
    \frac{(-1)^{\lambda }}{\sqrt{2^\lambda 4\ell }}\\
    \jj{\ell}{\xi}{\ell+\xi-\lambda}{\ell}{\xi-\lambda}{-(\ell+\xi-\lambda)}\simeq
    \frac{(-1)^{\lambda }}{\sqrt{2 \ell }}
\end{align}
while if we assume $m_1,m_2\lesssim \sqrt{\ell}$,
\begin{equation}
\label{eq:3j_general_app}
    \jj{\ell}{\ell-\xi}{2\ell-\xi-\lambda}{m_1}{m_2}{m}\simeq
    \frac{(-1)^{\xi+m}}{\ell^{3/4}\sqrt[4]{2\pi }}P_\lambda\left(\frac{m_1-m_2}{2\sqrt{\ell}}\right)e^{-\frac{(m_1-m_2)^2}{4\ell}}
\end{equation}
where $P_\lambda$ is a polynomial of degree $\lambda$, whose parity matches the parity of $\lambda$. For example
\begin{equation}
    P_0(x) = \frac{1}{\sqrt{2}},\;
    P_1(x) = \sqrt{2} x,\;
    P_2(x) = \frac{4 x^2-1}{2 }
\end{equation}
Lastly, for $\ell\times 2\rightarrow\ell+2$ we get
\begin{gather}
    \jj{\ell}{2}{\ell+2}{m_1}{m_2}{m} \simeq \\
    \frac{(-1)^{\ell +m}\sqrt{3}}{2\sqrt{ \ell  (2-m_2)! (m_2+2)!}}\left(1-\frac{m_1}{\ell }\right)^{1-\frac{m_2}{2}} \left(1+\frac{m_1}{\ell }\right)^{1+\frac{m_2}{2}}
\end{gather}
uniformly in $|m_1|\le\ell$, $|m_2|\le2$.

These expressions can be straightforwardly obtained by expanding exact expressions (using Stirling formula). For eq.~(\ref{eq:3j_general_app}) it is important to keep $\frac{m_i}{\sqrt{\ell}}$ fixed to obtain the correct expression.


\end{document}